\documentclass[onecolumn]{article}
\usepackage[vmargin={2cm,2cm},hmargin=2cm]{geometry}
\usepackage{graphicx}
\usepackage{amsmath,amssymb,cite,algorithm,algpseudocode,color,url}

\begin{document}

\title{Stochastic Power Processing through Logic Operation\\of Power Packets%
\thanks{This is a preprint of an article submitted by the authors for possible publication in Royal Society Open Science. }
}

\author{Shiu Mochiyama$^\dag$ \and
        Takashi Hikihara
}

\date{Kyoto University Katsura, Nishikyo, Kyoto, 615-8510 Japan \\[3mm]
        $^\dag$s-mochiyama@dove.kuee.kyoto-u.ac.jp
}

\maketitle

\begin{abstract}
    This article presents an application of the recently proposed logic operation of power based on power packetization. 
    In a power packet dispatching system, the power supply can be considered as a sequence of power pulses, where the occurrence of pulses follows a probability that corresponds to the capacity of the power sources or power lines. 
    In this study, we propose a processing scheme to reshape a stream of power packets from such stochastic sequences to satisfy the load demand. 
    The proposed scheme is realized by extending the concept of stochastic computing to the power domain. 
    We demonstrate the operation of the proposed scheme through experiments and numerical simulations by implementing it as a function of a power packet router, which forms a power packet dispatching network. 
    The stochastic framework proposed in this study provides a new design foundation for low-power distribution networks as an embodiment of the close connection between the cyber and physical components. 
\end{abstract}

    \section{Introduction}
    This is an update to the authors' previous article \cite{Inagaki.etal-2021}.
    In the previous article, a logic operation was defined for pulse-based power processing.  
    In addition, as an application of unary logic operation, error correction was proposed to compensate for the inaccuracy of the power supply due to noise and dissipation.
    This article presents the application of the defined binary operation to stochastic processing of pulse power as a novel approach to power management on a relatively low power distribution network. 

    The concept of logic operation of power \cite{Inagaki.etal-2021} is based on power packetization \cite{Takuno.etal-2010}. 
    Power is transferred as a train of pulse-shaped power; this is analogous to packet communication systems, where information is divided into smaller units of delivery called payloads. 
    Additionally, each power pulse has an information tag attached as a voltage signal that presents information regarding its origin, destination, and so on, in a similar way to the header in an information packet. 
    The combined unit of transfer is defined as a power packet. 
    The power packet routers in a power distribution network receive and forward the incoming packets according to the tags \cite{Takahashi.etal-2015,Takahashi.etal-2019}. 
    The logic operation of power was developed as a function of such power routers to determine an output pulsed-power sequence based on the input counterparts \cite{Inagaki.etal-2021}.
    
    The power packets are handled using a time-division multiplexing (TDM) method. 
    Thus, the handshake between sources and loads via a network of routers can be distinguished without mixing up together. 
    This provides the most important feature in a cyber-physical system: the avoidance of the discrepancy of information and physical quantity \cite{Kim.Kumar-2012,Guo.etal-2017}.

    The TDM transfer of power pulses leads to the representation of power as a digital sequence of unit pulses. 
    The power supply, demand, and line capacity are then represented by the sequence of 1/0, which corresponds to the existence/non-existence of a power packet \cite{Takahashi.etal-2016a,Mochiyama.Hikihara-2019a}. 
    This digital handling of power is quite different from the conventional PWM technique \cite{holmes2003pulse}, where the power flow is regulated using inherently analog techniques. 
    
    Furthermore, when the time interval of a power packet is set sufficiently short, the amount of power supply is represented by the density of power packets. 
    Subsequently, the power is represented by the probability, namely the number of power packets supplied within a certain period. 
    
    Probabilistic representation has been common both in communication and power engineering. 
    In communication networks, uncertainty in packet traffic is analyzed stochastically \cite{Kumar.etal-2004}. 
    Probabilistic handling of power flow also provides a practical approach to managing uncertainty in power systems \cite{Borkowska-1974,Dopazo.etal-1975,Sasaki.etal-2018,Gelenbe.Ceran-2016,Gelenbe.Abdelrahman-2018}. 
    A typical application is the inclusion of unpredictable power sources such as wind generators, photovoltaics, and other energy harvesting techniques that are subject to large fluctuations due to environmental conditions. 
    
    This study proposes a novel digitized scheme for stochasticity inclusion into power management as a physical layer technology. 
    We propose an application of the binary logic operation to stochastic power processing, focusing on the digital and stochastic representation of the power packet supply.
    We introduce the concept of stochastic computing \cite{Gaines-1969,Alaghi.Hayes-2013}, which was proposed as an alternative to deterministic computing for data processing, into power processing.
    In the stochastic computing scheme, a numerical value is represented by the probability of 1's in a binary number train. 
    The arithmetic operation is then realized by performing a logic operation of each bit of binary trains. 
    We correspond the inherently digital computing scheme to the management of power packets.
    
    The proposed scheme enables the constitution of an upstream allocation of load demand to power sources \cite{Bialek-1996} in a fully digitized and stochastic manner. 
    The packetized approach, unlike the conventional approaches involving continuous flow handling, can perform power management by accommodating uncertainty while retaining the most important feature of the system, namely the avoidance of the discrepancy in information and power handling. 
    
    The main contribution of this article can be summarized as follows. 
    We define stochastic power processing based on the logic operation of a power packet supply. 
    We then demonstrate the feasibility of the concept through two operations: multiplication and addition. 
    Furthermore, we apply this concept to power management between subnetworks in a power packet dispatching system. 
    We provide a strategy to fulfill the load's demand by coordinating the external and internal power supply, assuming that a subnetwork supplies its redundant power packets to another in a certain probability. 
    
    \section{Power packet dispatching system}
    Packetization of electricity was proposed in the 1990s as a method of power flow routing \cite{Toyoda.Saitoh-1998}, which was unfortunately not realized due to the limited technologies in the fields of power electronics, energy storage, and communication at the time. 
    Subsequently, following the development of the fundamental technologies, extensive research has been conducted on this concept based on both the theoretical and experimental aspects by multiple independent groups worldwide \cite{He.etal-2008,Takuno.etal-2010,Stalling.etal-2012,Eaves-2012,Gelenbe.Ceran-2016,Sugiyama.etal-2017}. 
    This study particularly focuses on the power packet dispatching system \cite{Takuno.etal-2010,Takahashi.etal-2015}, which helps in the physical realization of power packetization and routing, as a key enabler of power processing.
    
    \begin{figure}
        \centering
        \includegraphics[width=\linewidth]{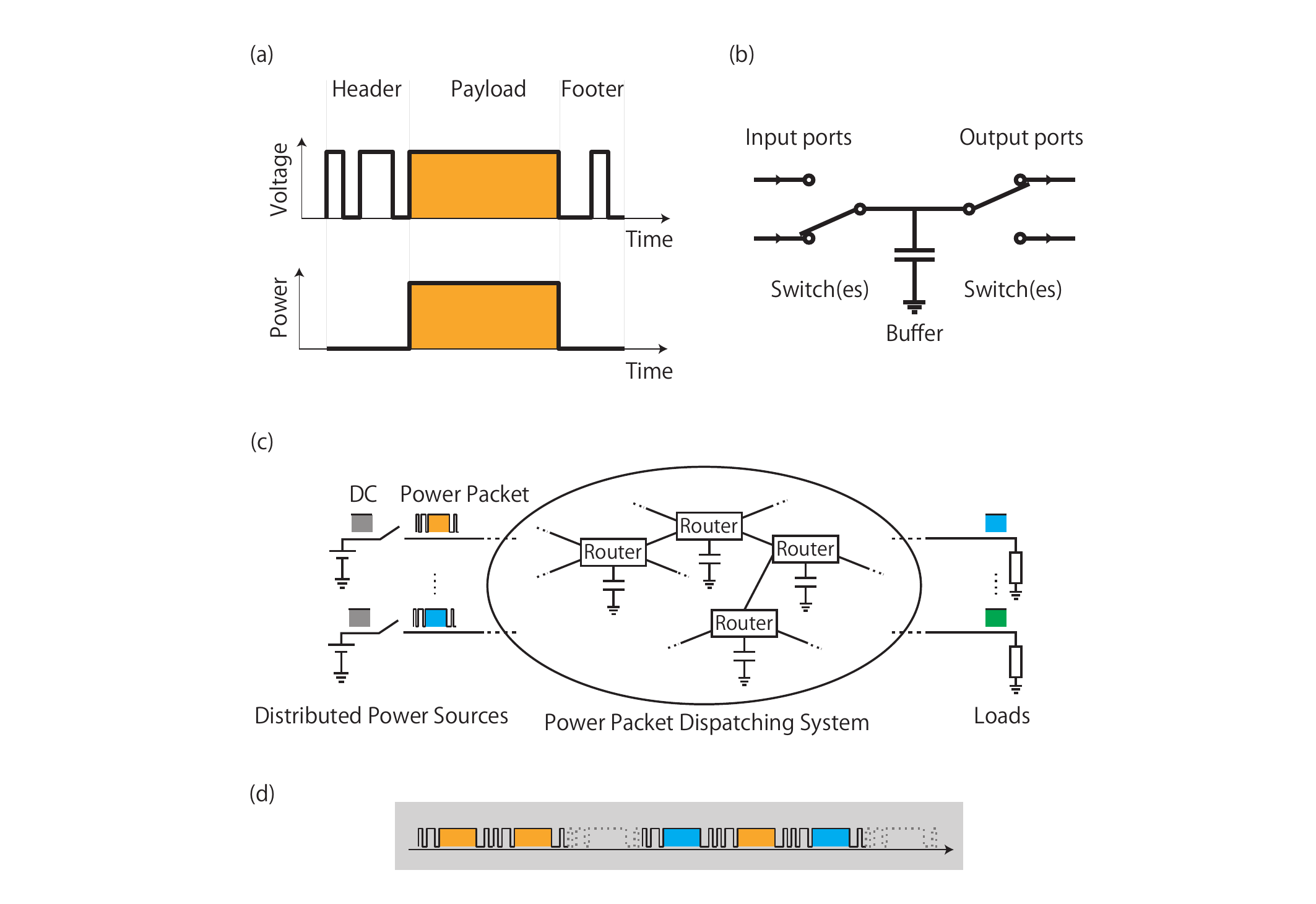}
        \caption{Power packetization and its dispatching system: (a) configuration of a power packet \cite{Takuno.etal-2010}, (b) circuit diagram of a router \cite{Takahashi.etal-2015,Takahashi.etal-2019}, (c) overview of a power packet dispatching system \cite{Takuno.etal-2010,Takahashi.etal-2015} and (d) example of density representation of power supply through time-division multiplexing \cite{Takahashi.etal-2016a,Mochiyama.Hikihara-2019a}. }
        \label{fig:packet}
    \end{figure}
    
    Figure~\ref{fig:packet} outlines the concept of power packetization. 
    Figure~\ref{fig:packet}~\textit{a} depicts the configuration of a power packet. 
    A payload is a unit of power transfer represented as a sequence of power pulses. 
    An information tag is then attached to the payload as a voltage waveform. 
    The tags before and after a payload are called a header and a footer. 
    For example, the header contains the destination address of the following payload. 
    An information tag is not accompanied by a current to avoid unnecessary power consumption. 
    
    Figure~\ref{fig:packet}~\textit{b} presents the conceptual circuit diagram of a power packet router. 
    The circuit comprises semiconductor switches that determine the flow of the payload based on the tag information. 
    A storage element is placed between the input and output ports to enable a router to buffer a power packet before forwarding. 
    
    Figure~\ref{fig:packet}~\textit{c} depicts an example of a power packet dispatching system.
    The power is supplied from sources to loads through a network of routers. 
    The network can distinguish each power packet throughout the routing path due to the TDM nature and the physical tag attachment. 
    This enables the system to dynamically reconfigure, e.g., attach/detach a subsystem containing several routers, sources, and loads.
    
    From the perspective of a load, the supply becomes a digital sequence of power packets, as shown in Fig.~\ref{fig:packet}~\textit{d}. 
    The loads require a certain density of power supply according to their operation.
    The router placed just before the load is supposed to collect the power packets to output the supply sequence to satisfy the demand. 
    The stochastic scheme proposed in this article corresponds to this function of a router. 
    
    \section{Logic operation of power packet}
    
    \begin{figure}
        \centering
        \includegraphics[width=\linewidth]{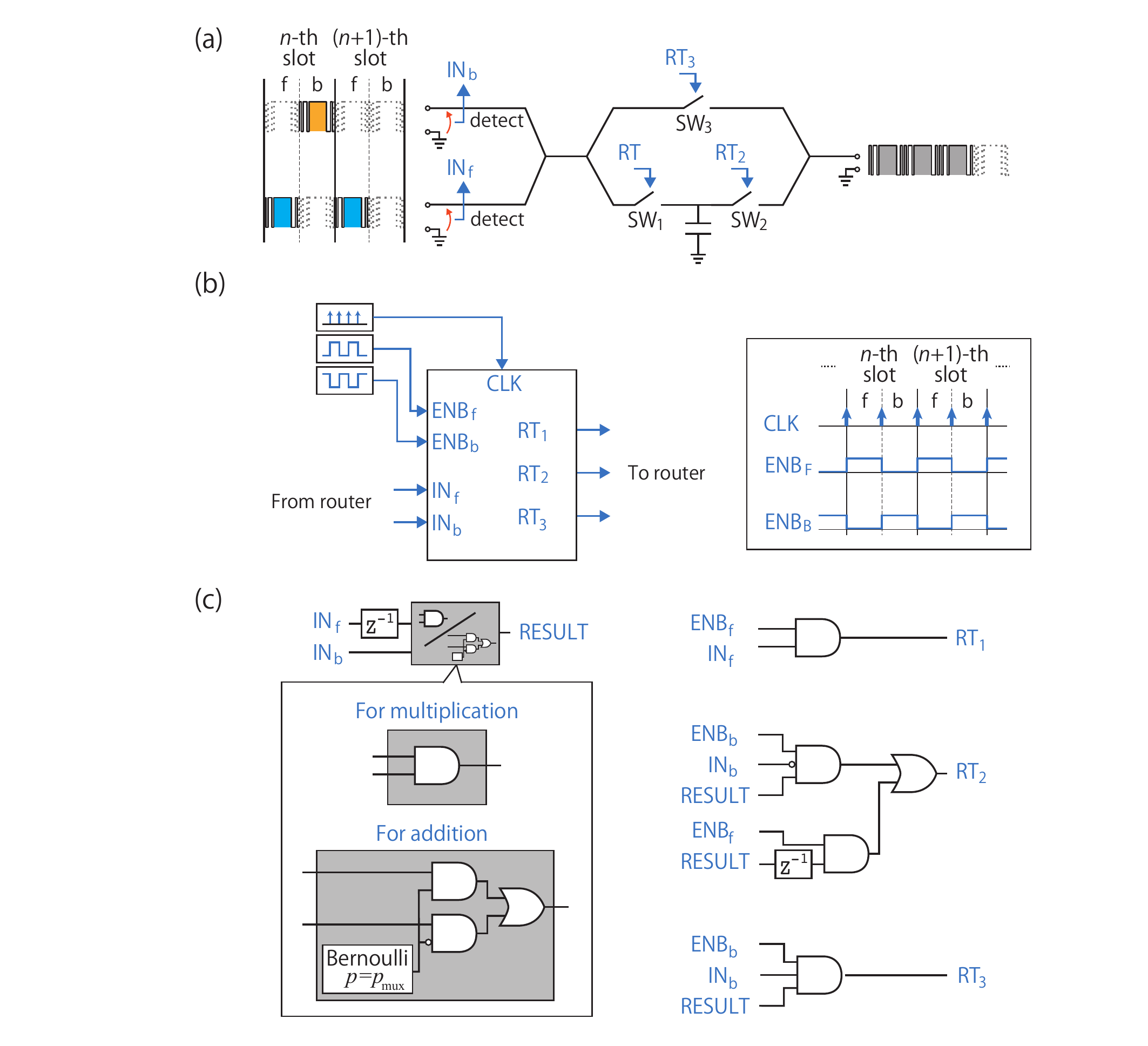}
        \caption{System configuration for logic operation of power packet: (a) router circuit, (b) in/out signals of the router's controller and (c) operation algorithm of the router's controller. }
        \label{fig:system}
    \end{figure}
    
    Logic operations are defined by assigning logic 1 and 0 corresponding to the presence and absence of power packets, respectively \cite{Inagaki.etal-2021}.
    The presence of a power packet is defined as its payload voltage being greater than a predetermined threshold value. 
    The threshold is set to a value that is sufficiently small but greater than noise levels.
    
    Figure~\ref{fig:system}~\textit{a} depicts the circuit configuration of a router for the logic operation. 
    The circuit is almost identical to what we used in our previous study \cite{Inagaki.etal-2021}. 
    The only difference is the position of the switch SW$_3$. 
    This change eliminates an on-loss of the semiconductor switch when the payload goes through the path that bypasses the buffer, presenting no essential difference in its essential operation.
    
    Following the setup in our previous study \cite{Inagaki.etal-2021}, logic operations are executed in synchronized time slots of a constant length.
    Figure~\ref{fig:system}~\textit{b} depicts the router controller which performs the logic operation by generating the gate signals. 
    For the binary operations%
    \footnote{We only consider binary operations in this article. For unary operations, execution occurs at every slot without distinction between f and b.}%
    , we assume that the slot comprises two consecutive intervals in a time-division manner. 
    These intervals are denoted as f and b, wherein f is followed by b.
    The outputs are the results of the operations of the logic of f and b.
    
    The software configuration has one major difference from the setup of our previous study. 
    In the previous setup, the router outputs a power packet only at the b interval since the result of the operation is not determined at the f interval yet. 
    From the perspective of output power regulation, this method limits the utilization of the output time intervals to half at most. 
    Since the amount of power is represented by the density of the pulse power, the regulation range of the output power is limited to $[0,0.5]$. 
    Therefore, the output will be quite small after passing through multiple operation circuits.
    To avoid this issue, we use both the f and b intervals for the output. 
    We define the output at the f interval as identical to that of the preceding b interval. 
    This configuration is possible since the router is equipped with a buffer. 
    The router can use the stored power for the output even when it has no incoming power packet at the f interval. 
    
    Figure~\ref{fig:system}~\textit{c} presents the details of the operation inside the router controller. 
    The operation of the two inputs, or the map from $\mathrm{IN}_\mathrm{f}$, $\mathrm{IN}_\mathrm{b}$ to $\mathrm{RESULT}$, is conducted based on the stochastic logic operation. 
    Their details will be provided later. 
    The map from the operation result to the switch signals is also represented by a combination of signal logic operations. 
    They are fixed for all operations and are implemented in the router controller's software. 
    
    \section{Stochastic Computing}
    In stochastic computing, a number is represented by a bitstream generated by a random process, i.e. a Bernoulli sequence.
    The value of a number is then defined by the number of 1s in the bitstream. 
    There are two major definitions for the map between a bitstream and a number: unipolar and bipolar configuration. 
    The former maps the probability to a number in $[0,1]$, while the latter maps the probability in $[-1,1]$. 
    In this study, we will consider the unipolar configuration since it is easy to let the number correspond to density-based power operation.
    Here, the value corresponds to the probability of the existence of 1's, that is the number of 1's divided by the bit length. 
    For example, a bitstream of length $N = 4$, "1010," corresponds to $0.5$. 
    The representation of a number is not unique; for example, $1100$, $1010$, and $101100$ are all interpreted as $0.5$. 
    
    For such stochastically defined numbers, arithmetic operations are described by bit-wise logic operations \cite{Gaines-1969,Alaghi.Hayes-2013}. 
    In this study, we will consider multiplication and addition of two numbers as examples.
    
    The multiplication is computed by performing the bit-wise AND operation of two stochastic numbers.
    Assuming that the probabilities of 1s in the two input numbers are $p_1$ and $p_2$, the output stochastic number is represented by $p_1 \cdot p_2$. 
    
    The addition is computed by applying a multiplexer with a selection variable as a stochastic number of probability $p_\mathrm{mux}$. 
    We then define the multiplexer output as follows; 
    output the logic of the first input if the selection variable is 1, otherwise output the logic of the second input. 
    The resulting output is represented by $p_\mathrm{in1} \cdot p_\mathrm{mux} + p_\mathrm{in2} \cdot (1-p_\mathrm{mux})$. 
    The addition is weighted by the selection variable. 
    We consider the equally weighted case, $p_\mathrm{mux} = 1/2$, throughout the article. 
    
    \section{Power processing based on stochastic operation}
    As stated earlier, we introduce the stochastic computing framework to packet-oriented power processing.
    The principal definition is quite straightforward. 
    The 1/0 in a stochastic number corresponds to the existence/nonexistence of a power packet. 
    
    The power source is considered as a random power packet generator that produces a power packet of unit power, $P_\mathrm{unit}$, at each time slot according to a Bernoulli distribution of 1's probability $p \in [0,1]$.
    The probability $p$ represents the availability of the power source in terms of the averaged time scale. 
    Power packet sequences are then input to a router, where the stochastic operation is performed on the sequence. 
    The resulting output is also a power packet sequence.
    
    The decoding is then performed at a load as power consumption.
    We consider an averaging operation as the decoding scheme of the received power. 
    Here, we assume that the window size of the average is sufficiently wide when compared to the power packet length, and simultaneously narrow enough to represent the changing power demand of the load%
    \footnote{The existence of such an appropriate window size can be ensured in several applications; a typical example is electromechanical actuation, where the electrical time constant is much smaller than the mechanical time constant related to the overall target operation. }. 
    Suppose that the resulting power packet sequence delivered to a load has 1's in the probability of $p'$ after the bit-wise logic operation based on the stochastic operation.
    Then, the average power consumption at the load would be $p' \cdot P_\mathrm{unit}$.
    
    In this study, we consider multiplication and addition as the operations performed at a router. 
    Figure~\ref{fig:system}~\textit{c} depicts the router's control scheme for the multiplication and addition operations. 
    We present the detailed configuration for the stochastic processing of power packets below. 
    
    \subsection{Multiplication}
    The multiplication is defined by an AND operation of the power packet inputs at each time slot. 
    The control signals of the router's switches are generated based on the result of the multiplication operation. 
    
    When the probability of two input sequences are $p_\mathrm{in1}$,$p_\mathrm{in2}$, the resulting output power is described as
    \begin{equation}
     P_\mathrm{out} = P_\mathrm{unit} \cdot p_\mathrm{in1} \cdot p_\mathrm{in2}.
     \label{eq:mul_def} 
    \end{equation}
    
    The operation can be extended to operations of three or more input streams by applying the AND operation with the corresponding number of inputs. 
    
    \subsection{Addition}
    The addition operation is realized by a multiplexer. 
    Here, in the same manner as in the stochastic computing, we introduce a stochastic selection variable of probability $p_\mathrm{mux}$. 
    The output of the multiplexer coincides with its first input if the selection variable is high and with its second input otherwise. 
    The overall operation is represented by a combination of logic gates as shown in Fig.~\ref{fig:system}.
    
    The output power is then described as
    \begin{equation}
     P_\mathrm{out} = P_\mathrm{unit} \cdot \left\{ p_\mathrm{in1} \cdot p_\mathrm{mux} + p_\mathrm{in2} \cdot (1-p_\mathrm{mux}) \right\}.
    \end{equation}
    Since we assume $p_\mathrm{mux} = 1/2$, we obtain
    \begin{equation}
     P_\mathrm{out} = P_\mathrm{unit} \cdot \frac{p_\mathrm{in1} + p_\mathrm{in2}}{2}. 
     \label{eq:add_def}
    \end{equation}
    
    The stochastic selection variable can be either a power packet input or an internal (signal) state.
    In this study, we set $p_\mathrm{mux}$ as an internal (virtual) variable for the sake of simplicity in controller design.
    
    Similar to the multiplication, the addition can be extended to operations of three or more input streams by increasing the inputs of the multiplexer. For the addition, the selection variable must also be extended to the corresponding bit width. 
    
    \subsection{Closure of packetized power expression under stochastic operations}
    Since the operations are implemented as a function of a power router, they may be performed several times during the power routing from a source to a load with multiple hops. 
    This requires the packetized expression of power to be closed under the defined operations. 
    
    In the definition of stochastic power processing, the (normalized) power supply is expressed as a probability $p\in [0,1]$. 
    Every value of power that can be considered is closed under the defined multiplication and addition. 
    That is, for any $p_1,p_2\in [0,1]$, the results of opeartions $p_\mathrm{mul} = p_1 p_2$ and $p_\mathrm{add} = p_1 p_\mathrm{mux} + p_2 (1-p_\mathrm{mux})$ satisfy $p_\mathrm{mul}, p_\mathrm{add} \in [0,1]$. 
    
    Furthermore, the closure is confirmed for an instantaneous power in/out relationship. 
    The instantaneous power output is obtained as a result of logic operations. 
    Consider $s_1, s_2 \in \{0,1\}$ as the variables that correspond to the absence or presence of input power packets at a unit time interval and $s_\mathrm{mux} \in \{0,1\}$ as a logical variable for a multiplexer input. Then, the outputs of multiplication and addition are defined as $s_1 \land s_2$ and $(s_1 \land s_\mathrm{mux}) \lor (s_2 \land \neg s_\mathrm{mux})$. 
    These two results are also the logical values of $\{0,1\}$.
    
    Based on the discussions above, we conclude that the proposed operations can be applied an arbitrary number of times along a routing path. 
    
    \section{Verification of Stochastic Operation of Power}
    \subsection{Setups}
    
    \begin{figure}
        \centering
        \includegraphics[width=\linewidth]{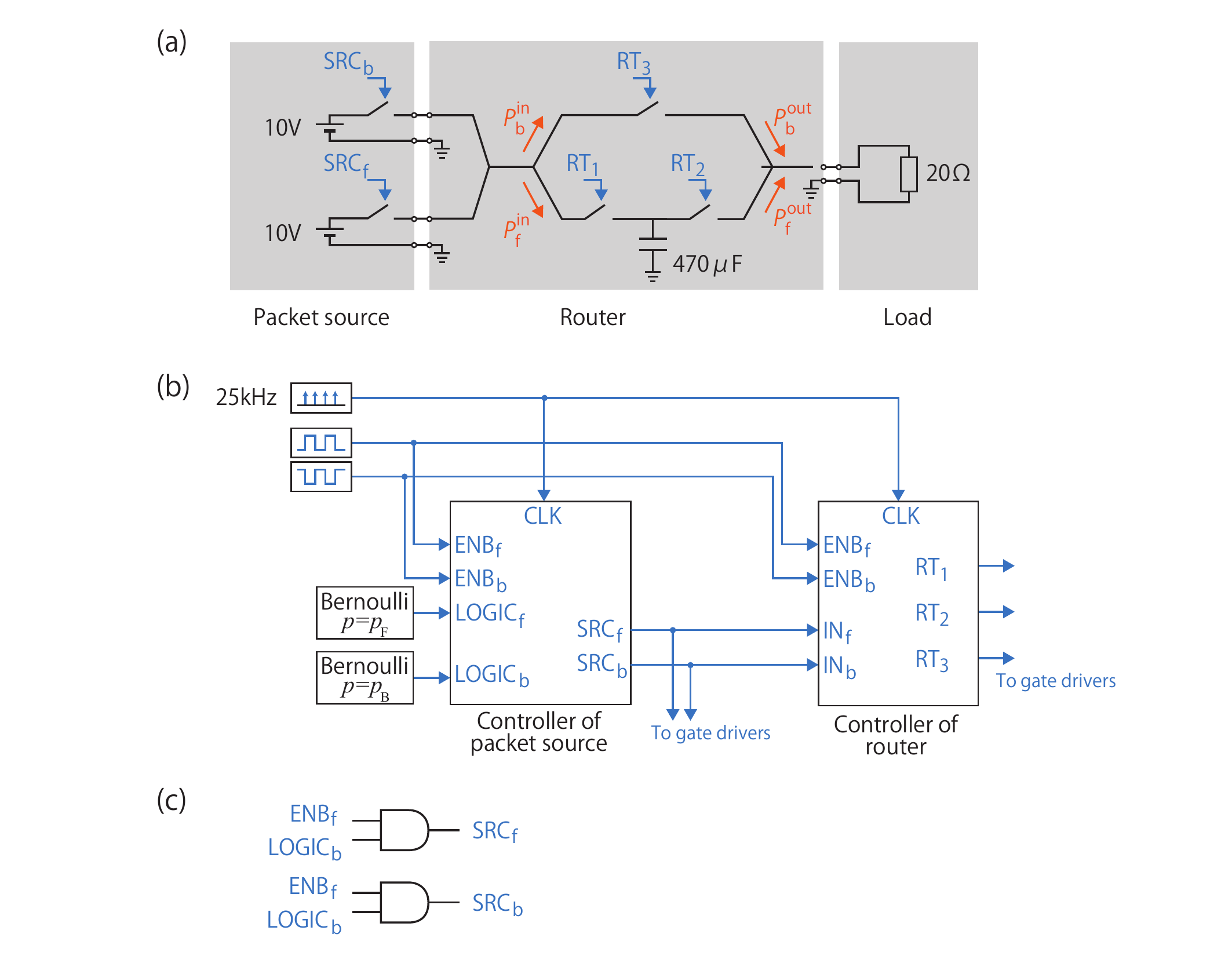}
        \caption{Setups for experimental verification: (a) circuit configuration, (b) in/out signals of the controllers of the packet source and the router, and (c) operation algorithm of the packet source. }
        \label{fig:setup}
    \end{figure}
    
    Figure~\ref{fig:setup} depicts the setups for the experimental verification. 
    We consider a system of two packet sources: one router with stochastic logic operation function and one load. 
    
    Figure~\ref{fig:setup}~\textit{a} depicts the circuit configuration of the whole system.
    The input of the router circuit comprises the two power packet sources. 
    They produce power packets of $10\,\mathrm{V}$ at different probabilities. 
    The gate signals of the switches at the two input ports are denoted as $\mathrm{SRC}_\mathrm{f}$ and $\mathrm{SRC}_\mathrm{b}$, indicating that they are in charge of intervals f and b, respectively. 
    Similarly, the probabilities of packet presence at the input ports are denoted by $p_\mathrm{f}$ and $p_\mathrm{b}$, respectively. 
    For the switches, we use bidirectional switching units comprising two back-to-back MOSFETs. 
    For the MOSFETs, we adopt part number SH8KB7TB1 from ROHM CO., LTD. 
    The load is set at $20\,\Omega$. 
    
    The gate signals of the switches in the packet sources and the router are controlled by clock-synchronized controllers, as shown in Fig.~\ref{fig:setup}~\textit{b}. 
    The clock frequency of the controller is set at $25\,\mathrm{kHz}$, which corresponds to the packet length of $40\,\mu s$. 
    
    Inside the controller of the packet sources, the gate signals are controlled based on the f/b slot division and Bernoulli distribution of $p_\mathrm{f}$ and $p_\mathrm{b}$, as shown in Fig.~\ref{fig:setup}~\textit{c}. 
    Please refer to Fig.~\ref{fig:system}~\textit{c} for the details of the gate signal control in the controller of the router. 
    
    We set up multiple cases for the combination of $p_\mathrm{f}$, $p_\mathrm{b}$, and the operation to be applied (multiplication or addition) for the verification. 
    Table~\ref{tab:setup} presents the list of setups.
    \begin{table}
        \caption{Probability setups for verification. }
        \label{tab:setup}
        \centering
        \begin{tabular}{cccc}
            \hline
            Case index & Operation & $p_\mathrm{f}$ & $p_\mathrm{b}$ \\ \hline 
            0 & Multiplication & 0.9 & 0.9 \\ 
            1 & Multiplication & 0.8 & 0.9 \\ 
            2 & Multiplication & 0.7 & 0.8 \\ 
            3 & Multiplication & 0.5 & 0.8 \\ 
            4 & Multiplication & 0.4 & 0.5 \\ 
            5 & Multiplication & 0.2 & 0.9 \\ 
            6 & Multiplication & 0.9 & 0.2 \\ 
            7 & Multiplication & 0.8 & 0.5 \\ \hline
            8 & Addition       & 0.9 & 0.9 \\ 
            9 & Addition       & 0.8 & 0.9 \\ 
           10 & Addition       & 0.7 & 0.8 \\ 
           11 & Addition       & 0.5 & 0.8 \\ 
           12 & Addition       & 0.4 & 0.5 \\ 
           13 & Addition       & 0.2 & 0.9 \\ 
           14 & Addition       & 0.9 & 0.2 \\ 
           15 & Addition       & 0.8 & 0.5 \\ \hline
        \end{tabular}
    \end{table}
    We fix the seeds of the random generators for each case listed in the table. 
    We will analyze the effects of different random seeds in the later section. 
    
    As discussed earlier, the result of stochastic power processing is evaluated by taking an average of the power consumption at the load during a $400\,\mathrm{ms}$ interval. 
    Additionally, some of the figures in the following sections present the output power in a normalized value. 
    The base value for the normalization is obtained by measuring the output power with $(p_\mathrm{f},p_\mathrm{b})=(1,1)$, i.e. all the switches are kept on. 
    
    The router circuit contains two paths of power transfer: one going through $\mathrm{SW}_1$ and $\mathrm{SW}_2$, and the other going through $\mathrm{SW}_3$.
    We denote these paths as "path f" and "path b", respectively. 
    The trace of power input/output is measured separately for the two paths. 
    We denote the measurements by orange arrows, e.g. input power for the path f by $P^\mathrm{in}_\mathrm{f}$, as shown in Fig.~\ref{fig:setup}.
    
    In the following experiments, the tag attachment and its reading are omitted in the procedure of power packet transfer.
    That is, the signals $\mathrm{SRC}_\mathrm{x}$ and $\mathrm{IN}_\mathrm{x}$ ($\mathrm{x}=\{\mathrm{f},\mathrm{b}\}$) are forced to be identical in the software layer. 
    This is done to simplify the design of the router's controller to focus on tracing the power flow. 
    The simplification is justified since the software/hardware configurations for these procedures are already established and validated experimentally \cite{Takahashi.etal-2015,Tanaka.etal-2020,Inagaki.etal-2021}. 
    
    \subsection{Results of multiplication}
    \begin{figure}
        \centering
        \includegraphics[width=\linewidth]{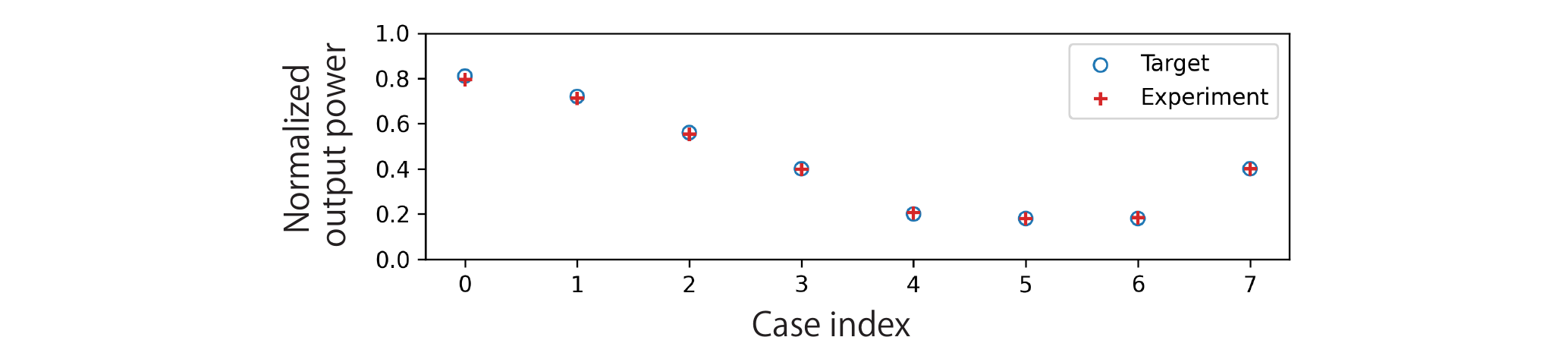}
        \caption{Result of multiplication: average power consumption at each case. }
        \label{fig:result_mul_ave}
    \end{figure}
    \begin{figure}
        \centering
        \includegraphics[width=\linewidth]{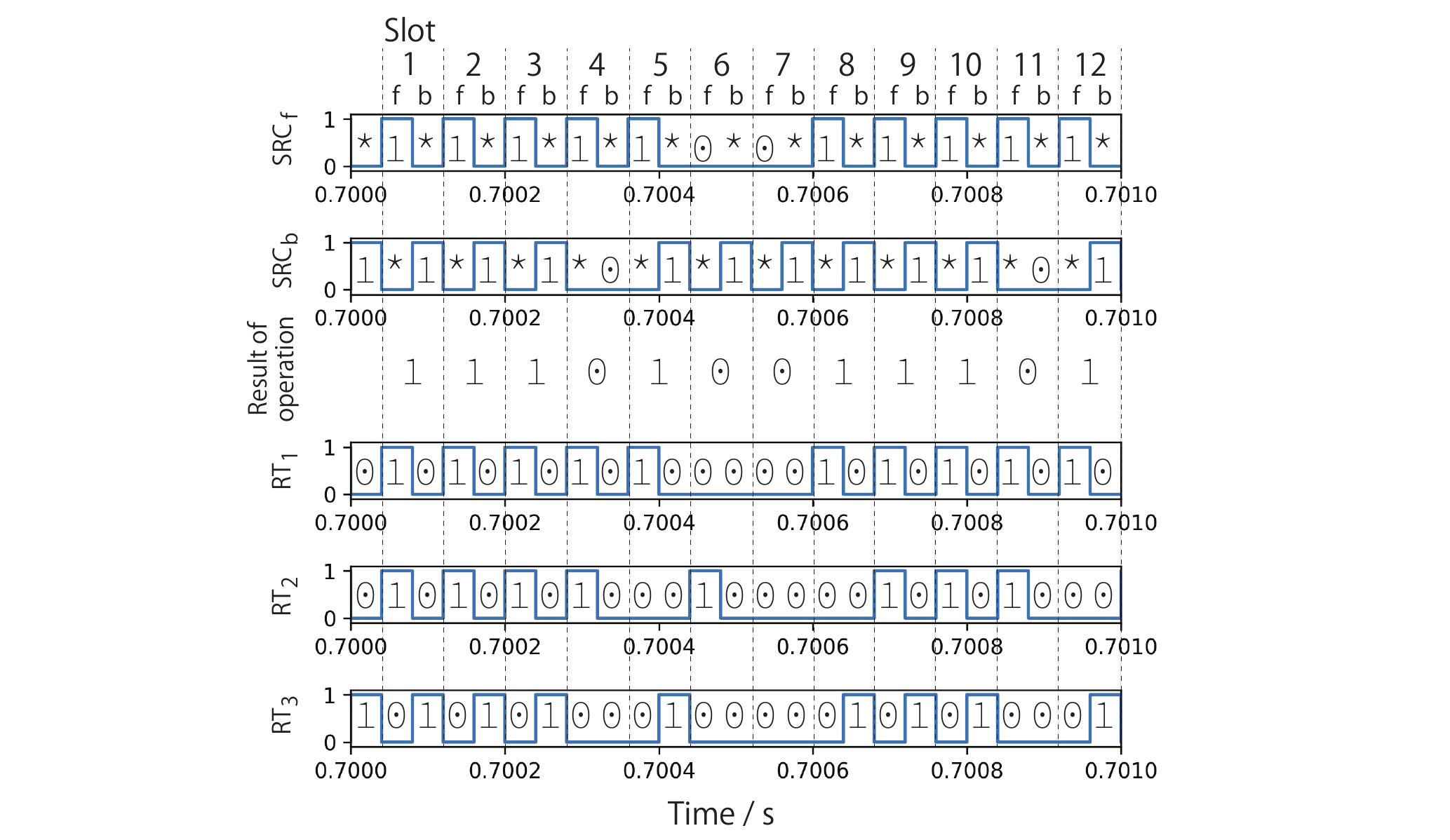}
        \caption{Result of multiplication of case 1: control signals of the packet sources and the router. }
        \label{fig:result_mul_sig}
    \end{figure}
    \begin{figure}
        \centering
        \includegraphics[width=\linewidth]{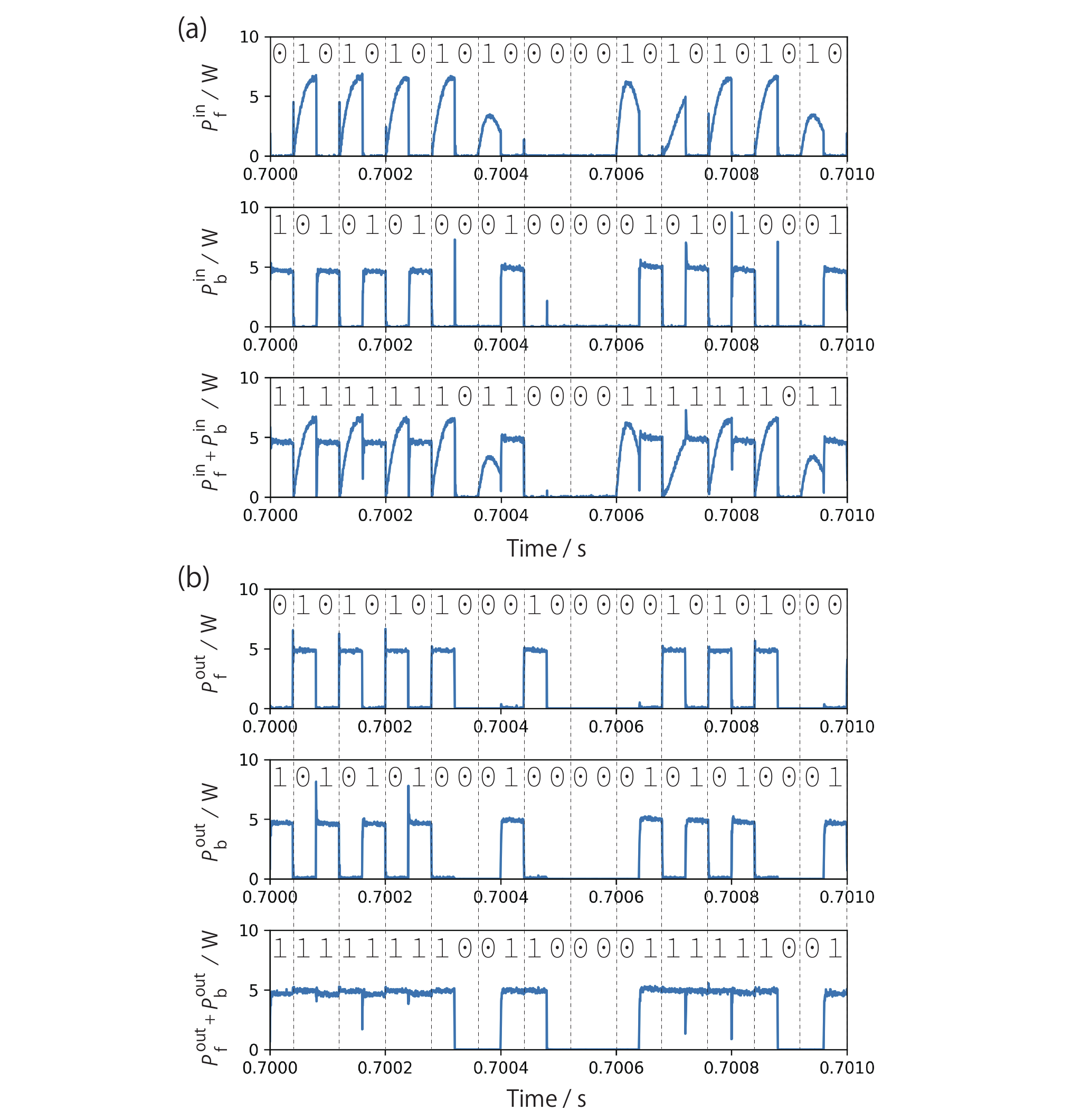}
        \caption{Result of multiplication of case 1: (a) input power flow and (b) output power flow. }
        \label{fig:result_mul_pow}
    \end{figure}
    
    In this subsection, we consider the results of the experiments of the multiplication operation. 
    First, we observe the result of power processing as an average supply to the load. 
    Figure~\ref{fig:result_mul_ave} presents the theoretical target value and the measured value of the average power supply to the load. 
    The measured power concurs well with the theoretical value for all the cases. 
    
    We then present the detailed results in one of the setup combinations of $(p_\mathrm{f},p_\mathrm{b})$.
    Figures~\ref{fig:result_mul_sig}~and~\ref{fig:result_mul_pow} depict the signals processed in the router's controller and the power trace measurements for case 1. 
    
    Figure~\ref{fig:result_mul_sig} shows that the control signals for the router switches are appropriately generated corresponding to the input power packet sequences and the stochastic logic operation scheme. 
    For a detailed explanation of the operation of the router, we consider time slot 2, where the inputs are both 1. 
    The presence of a power packet at the f interval triggers $\mathrm{RT}_1 = 1$, indicating that the input power is stored at the buffer. 
    The presence of a power packet at the b interval then yields 1 for the operation result of the slot. 
    Since the resulting output is 1 and the logic at the b interval is also 1, the output power is obtained through path b. 
    That is, $\mathrm{RT}_3 = 1$ holds at the b interval. 
    We can also confirm that the router's control signals were appropriately generated for all the other combinations of $(\mathrm{IN}_1,\mathrm{IN}_2)$ in the other time slots.
    
    Figures~\ref{fig:result_mul_pow}~\textit{a}~and~\textit{b} demonstrate that the power input/output is appropriately managed according to the result of logic operations, i.e. to the gate signals $\mathrm{RT}_i$ ($i=1,2,3$).
    We verify this by considering time slot 2 as an example again. 
    The input through path f, $P^\mathrm{in}_\mathrm{f}$, exists when $\mathrm{RT}_1 = 1$ holds. 
    Here, the shapes of the pulsed power waveforms differ in each time slot. 
    This is because the buffer's state of charge at the beginning of each f interval is not identical but depends on the history of the input logic sequences. 
    The power inflow through path b, $P^\mathrm{in}_\mathrm{b}$, occurs when $\mathrm{RT}_3 = 1$ holds. 
    In this case, the power is transferred directly from the source to the load; thus, the shapes of the pulse waveforms are identical in the constant load setup of this study. 
    The power output from the buffer, $P^\mathrm{out}_\mathrm{f}$, is generated at two conditions. 
    One is that power is output at f intervals if the operation result of the previous slot is 1. 
    As explained earlier, this is introduced to ensure a balance between average power input/output. 
    The other condition is that power is output at b intervals if the operation result is 1 but the input at b is 0. 
    This condition is never satisfied in the multiplication mode, where the result becomes 1 only when the inputs of both intervals are 1. 
    This type of supply can occur in the addition mode. 
    In summary, the waveform of $P^\mathrm{out}_\mathrm{f}$ is supposed to coincide with the operation result of the previous time slot, which can be confirmed by the figure. 
    The direct output at interval b, $P^\mathrm{out}_\mathrm{b}$ is identical to $P^\mathrm{in}_\mathrm{b}$. 
    This is also because the output becomes 1 only when the input at b is 1. 
    The above analysis of the four power waveforms demonstrates the successful routing of the power packets in the multiplication mode.

    \subsection{Results of addition}
    
    \begin{figure}
        \centering
        \includegraphics[width=\linewidth]{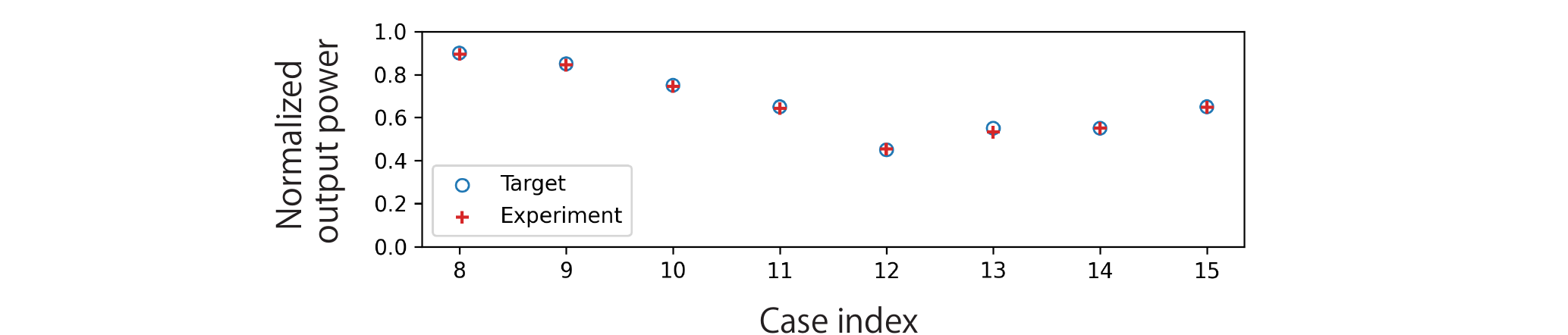}
        \caption{Result of addition: average power consumption at each case. }
        \label{fig:result_add_ave}
    \end{figure}
    \begin{figure}
        \centering
        \includegraphics[width=\linewidth]{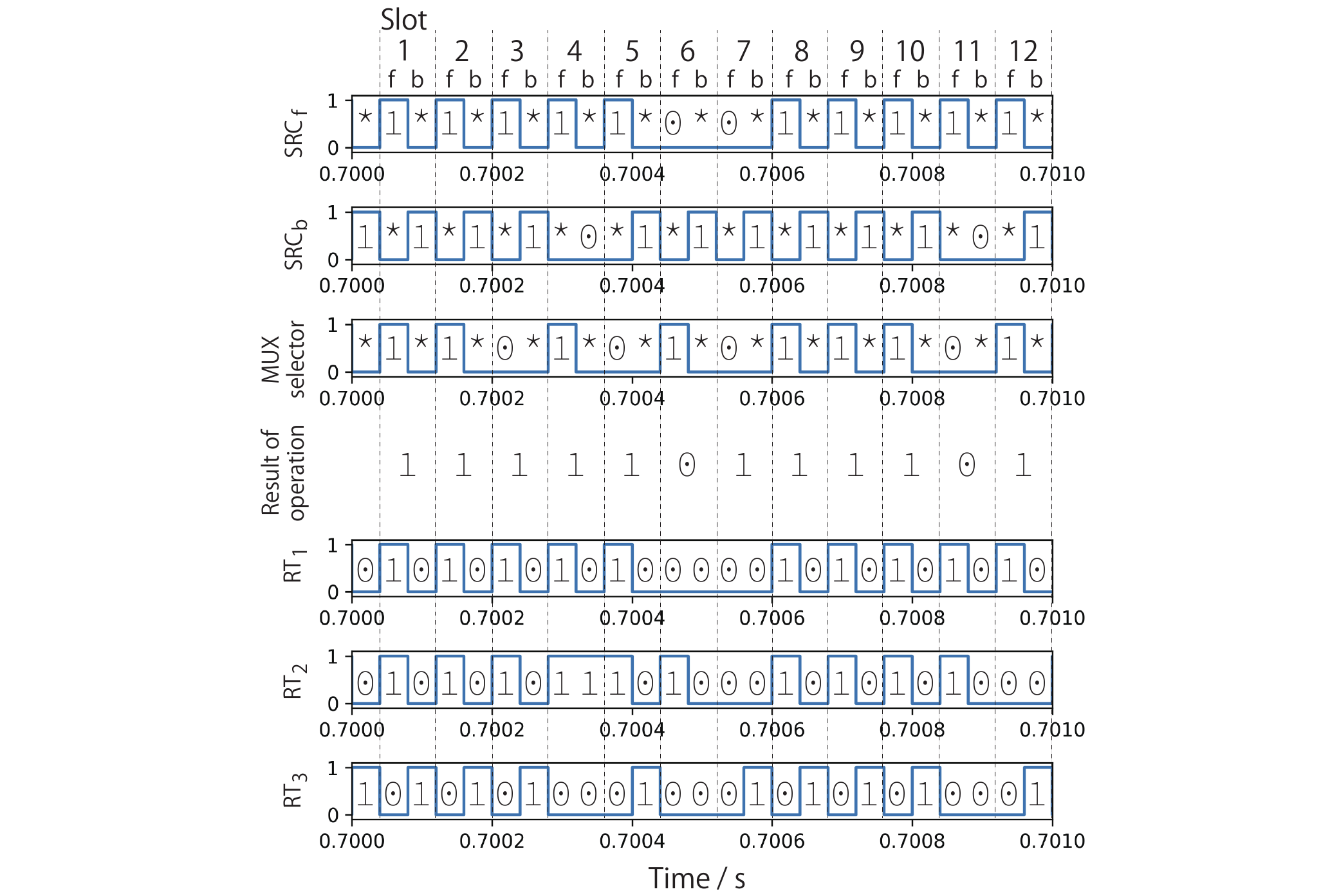}
        \caption{Result of addition of case 9: control signals of the packet sources and the router. }
        \label{fig:result_add_sig}
    \end{figure}
    \begin{figure}
        \centering
        \includegraphics[width=\linewidth]{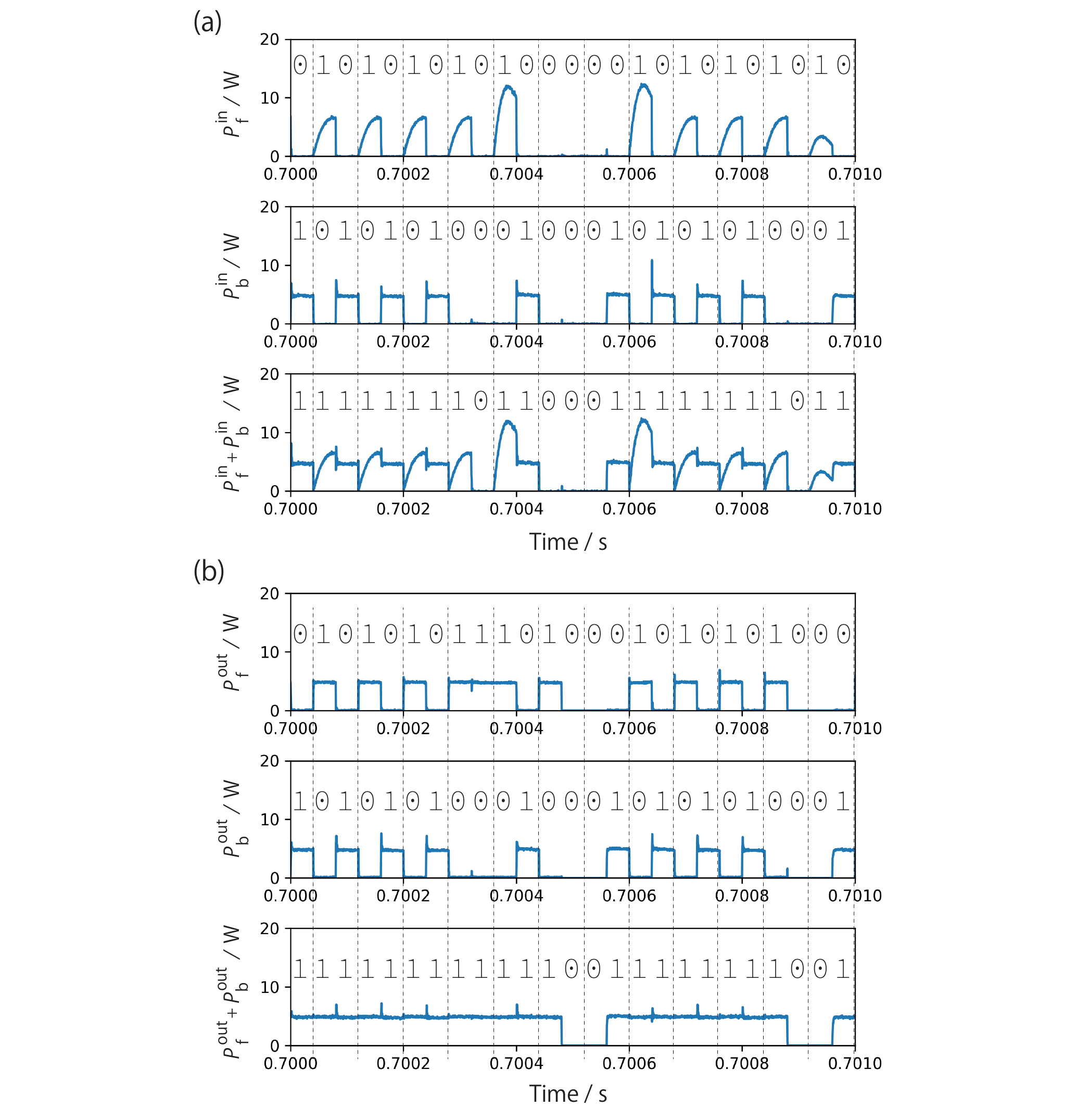}
        \caption{Result of addition of case 9: (a) input power flow and (b) output power flow. }
        \label{fig:result_add_pow}
    \end{figure}
    
    First, we consider the average supply to the load in the addition mode.
    The setups of the probabilities are identical to that in the multiplication mode's results, but the output is of course different. 
    Figure~\ref{fig:result_add_ave} depicts the theoretical target value and the measured value of the average power supply to the load. 
    Similar to the results of the multiplication mode, the measured power concurs well with the theoretical value for all the cases. 
    
    Subsequently, we present the detailed results for case 9, where $(p_\mathrm{f},p_\mathrm{b})=(0.8,0.9)$.
    Figures~\ref{fig:result_add_sig}~and~\ref{fig:result_add_pow} depict the signals processed in the router's controller and the power trace measurements. 
    
    Figure~\ref{fig:result_add_sig} demonstrates that the control signals for the router switches are appropriately generated corresponding to the input power packet sequences, the internal probabilistic variable given to the multiplexer, and the stochastic logic operation scheme. 
    We now analyze time slot 4, where the inputs are $(\mathrm{SRC}_\mathrm{f},\mathrm{SRC}_\mathrm{b})=(1,0)$ and $\mathrm{SRC}_\mathrm{mux}=1$. 
    The presence of a power packet at the f interval triggers $\mathrm{RT}_1 = 1$, indicating that the input power is stored at the buffer. 
    Then, $\mathrm{SRC}_\mathrm{mux} = 1$ indicates that the operation result is identical to $\mathrm{SRC}_\mathrm{f}$ regardless of the absence of a power packet at b. 
    That is, the output is 1 for the time slot. 
    Since the resulting output is 1 and the logic at the b interval is 0, the output power is obtained from the buffer through path f. 
    That is, $\mathrm{RT}_2$ is turned to 1 while $\mathrm{RT}_3$ is maintained at 0. 
    In the other time slots, we can also confirm that the router's control signals were appropriately generated for all the other combinations of $(\mathrm{IN}_1,\mathrm{IN}_2,\mathrm{SRC}_\mathrm{mux})$.
    
    Figures~\ref{fig:result_add_pow}~\textit{a}~and~\textit{b} demonstrate that the power input/output is appropriately managed based on the result of logic operations.
    We focus on time slot 4 again for further details. 
    The input through path f, $P^\mathrm{in}_\mathrm{f}$, exists at inverval f since $\mathrm{RT}_1 = 1$ holds. 
    Similar to the multiplication mode, the shapes of the pulsed power waveforms differ based on the history of the input sequences. 
    The power inflow through path b, $P^\mathrm{in}_\mathrm{b}$, does not occur since $\mathrm{SRC}_\mathrm{b} = 0$ holds. 
    Subsequently, the power output from the buffer, $P^\mathrm{out}_\mathrm{f}$, is obtained at both the f and b intervals. 
    The power output at interval f is generated since the operation result of the previous slot is 1. 
    The supply at interval b exists since the operation result is 1 but the input at b is 0. 
    The above results demonstrate the successful routing of power packets in the addition mode. 
    
    \subsection{Statistical properties of output power}
    Since the proposed scheme depends on the stochastic operation, the outcome also becomes stochastic. 
    In all of the aforementioned experiments, we analyzed only one sample with the same set of random seeds. 
    In this subsection, we analyze the statistical properties of the output power based on a larger number of samples with different seed setups.
    
    For the analysis, we perform 100 trials with different sets of seeds for each of the 16 cases shown in Table~\ref{tab:setup}. 
    Owing to the large number of trials involved, we perform numerical simulations instead of experiments. 
    Before analyzing the numerical results, we first confirm the validity of the numerical model through a comparison with the experimental results. 
    
    \subsubsection{Comparison of numerical and experimental results}
    The setup of the numerical model is basically identical to that in the experiment. 
    The only difference is that we use LCR components to represent the parasitics mainly due to the lines. 
    The values of inductance, capacitance, and resistance are set at $10\,\mathrm{nH}$, $100\,\mathrm{pF}$, and $0.2\,\mathrm{m}\Omega$, respectively. 
    
    \begin{figure}
        \centering
        \includegraphics[width=\linewidth]{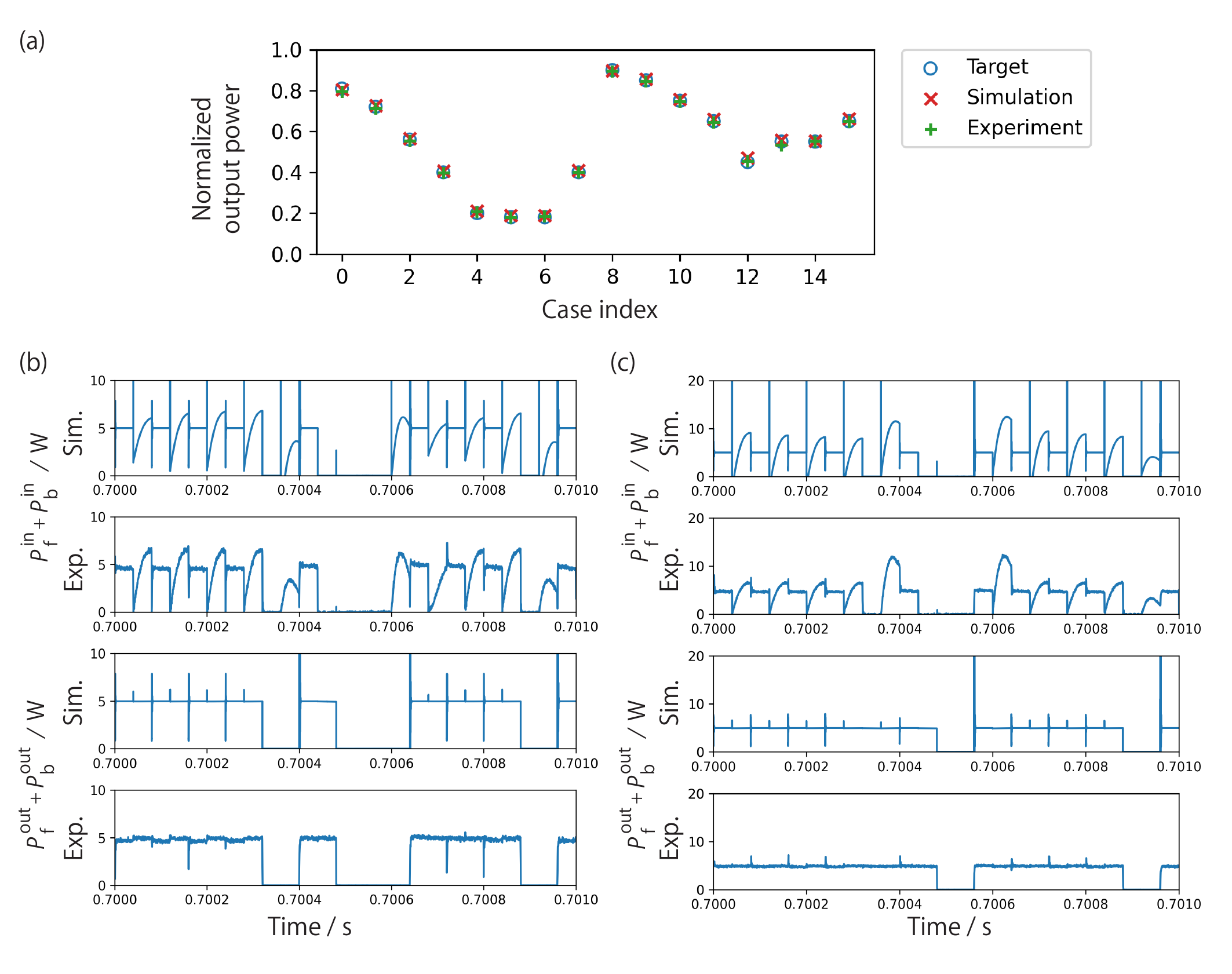}
        \caption{Comparison of results of the experiment and the numerical simulation: (a) average output power, (b) power in/out waveforms in case 1 and (c) power in/out waveforms in case 9. }
        \label{fig:sim_vs_exp}
    \end{figure}
    Figure~\ref{fig:sim_vs_exp} presents the comparison of the results obtained through the experiment and the numerical simulation.
    Figure~\ref{fig:sim_vs_exp}~\textit{a} presents the normalized output when compared to the experimental value. 
    The simulation values of the output power are normalized in the same way as the experiment. 
    The numerical results concur well with the experiment in terms of the average output. 
    Subsequently, we analyze the detailed power waveforms in cases 1 and 9. 
    Figures~\ref{fig:sim_vs_exp}~\textit{b}~and~\textit{c} depict the power inflow and outflow of cases 1 and 9, respectively. 
    Both the input and output power waveforms concur well with each other. 
    The only exception is the slight difference in the shape of the inflow to the buffer. 
    This is primarily attributed to the estimation error of the LC values of the parasitic elements of the circuit. 
    Still, Fig.~\ref{fig:sim_vs_exp} demonstrates that this error affects only the shape of the instantaneous waveform and not the average power supply. 
    Therefore, we retain the numerical model used here in the following investigations.
    
    \subsubsection{Statistical analysis}
    \begin{table}
        \caption{Statistical analysis of the output power with 200 samples for each case. The case index is the same as in Table~\ref{tab:setup}. }
        \label{tab:seed}
        \centering
        \begin{tabular}{cccc}
            \hline
            Case index & Mean & Unbiased variance & Statistic \\ \hline 
            0  & 0.809403 & 0.013607 & -0.072344  \\
            1  & 0.710417 & 0.017620 & -1.020960  \\
            2  & 0.559056 & 0.019541 & -0.095488  \\
            3  & 0.396618 & 0.019082 & -0.346233  \\
            4  & 0.201164 & 0.013604 &  0.141193  \\
            5  & 0.170072 & 0.012054 & -1.278801  \\
            6  & 0.180449 & 0.013470 &  0.054743  \\
            7  & 0.404319 & 0.018983 &  0.443348  \\
            8  & 0.903492 & 0.006513 &  0.611926  \\
            9  & 0.844969 & 0.010909 & -0.681160  \\
            10 & 0.751402 & 0.015852 &  0.157469  \\
            11 & 0.652395 & 0.020711 &  0.235314  \\
            12 & 0.451970 & 0.020485 &  0.194675  \\
            13 & 0.553703 & 0.022586 &  0.348485  \\
            14 & 0.547185 & 0.021318 & -0.272691  \\
            15 & 0.652706 & 0.016566 &  0.297365  \\  \hline
        \end{tabular}
    \end{table}
    Here, we focus on the statistical analysis. 
    We expect that the averaged output power is obtained from the normal distribution of mean value equal to the target value, i.e. $p_1p_2$ for multiplication and $(p_1+p_2)/2$ for addition, as a result of the stochastic power processing. 
    The averaging window for the output power calculation is defined as the $1\,\mathrm{ms}$ interval in this analysis. 
    We perform a two-sided t-test on the mean value of a population with unknown variance \cite{Schmetterer-1974} based on the samples taken with different random seed setups. 
    
    Thus, the hypothesis for the t-test on each of the 16 cases is that the data of $N=200$ samples is obtained from a population with a mean equal to its target value and an unknown variation. 
    The following statistic is defined to test the hypothesis:
    \begin{equation}
        t = \frac{\bar{x} - \mu}{\sqrt{\frac{s^2}{N}}},
    \end{equation}
    where $\bar{x}$, $s^2$, and $\mu$ represent the mean of the samples, unbiased estimate of variance, and population mean, respectively. 
    We set the significance level at 0.05, which is a commonly used value. 
    Based on the t-distribution's degree-of-freedom at $200-1=199$, the critical value becomes $\pm 1.972$. 
    
    Table~\ref{tab:seed} presents the mean, unbiased estimate of variance, and statistic. 
    The mean and variance values are normalized in the same way as in the previous experiments, i.e. calculated with the output power normalized by the maximum output with $(p_\mathrm{f},p_\mathrm{b}) = (1,1)$. 
    Thus, the population mean also becomes the normalized value i.e. $p_1p_2$ for multiplication and $(p_1+p_2)/2$ for addition. 
    Consequently, the values of the statistic in all the cases do not exceed the critical value.
    Thus, the test concluded that the hypothesis is accepted in all the cases. 
    
    \section{Stochastic power processing for power management}
    In this section, we demonstrate an example of power management using the proposed stochastic power processing. 
    We consider an interaction between independent subsystems, each of which manages its own power supply in a packetized manner. 
    We then focus on the power management based on the stochastic power processing in a subsystem by utilizing both its own source and a share of surplus power from another subsystem. 
    
    \begin{figure}
        \centering
        \includegraphics[width=\linewidth]{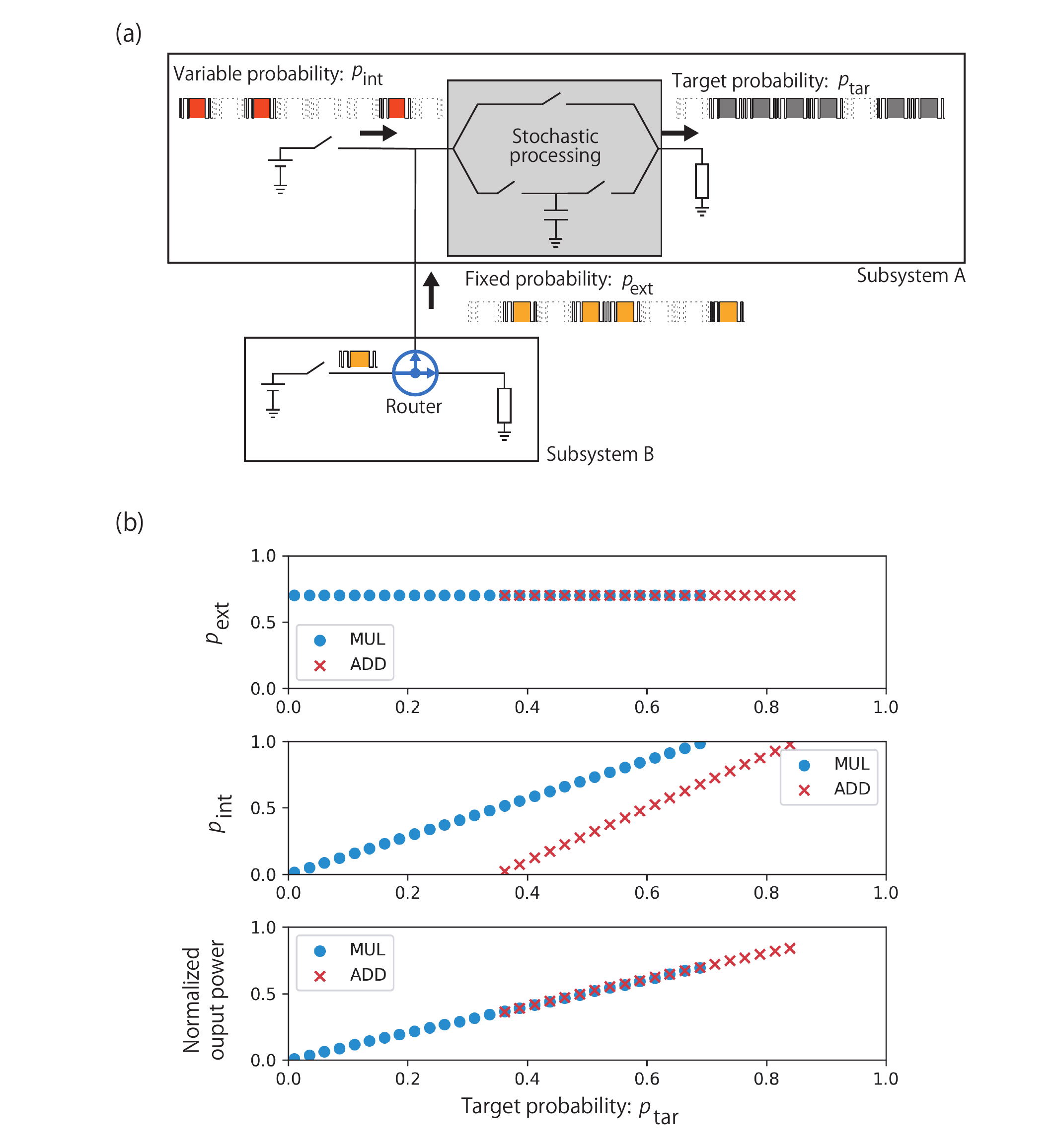}
        \caption{Overview of power management based on stochastic power processing: (a) system setup and (b) visualization of probability setup as a function of the target probability. In (b), the top graph shows that the probability of external input is fixed for any target value; the middle shows that the probability of variable input changes based on the relationship (\ref{eq:mul}) and (\ref{eq:add}); the bottom graph shows that the probability of the overall output follows the relationship (\ref{eq:mul_def}) and (\ref{eq:add_def}). }
        \label{fig:manage1}
    \end{figure}
    \begin{figure}
        \centering
        \includegraphics[width=\linewidth]{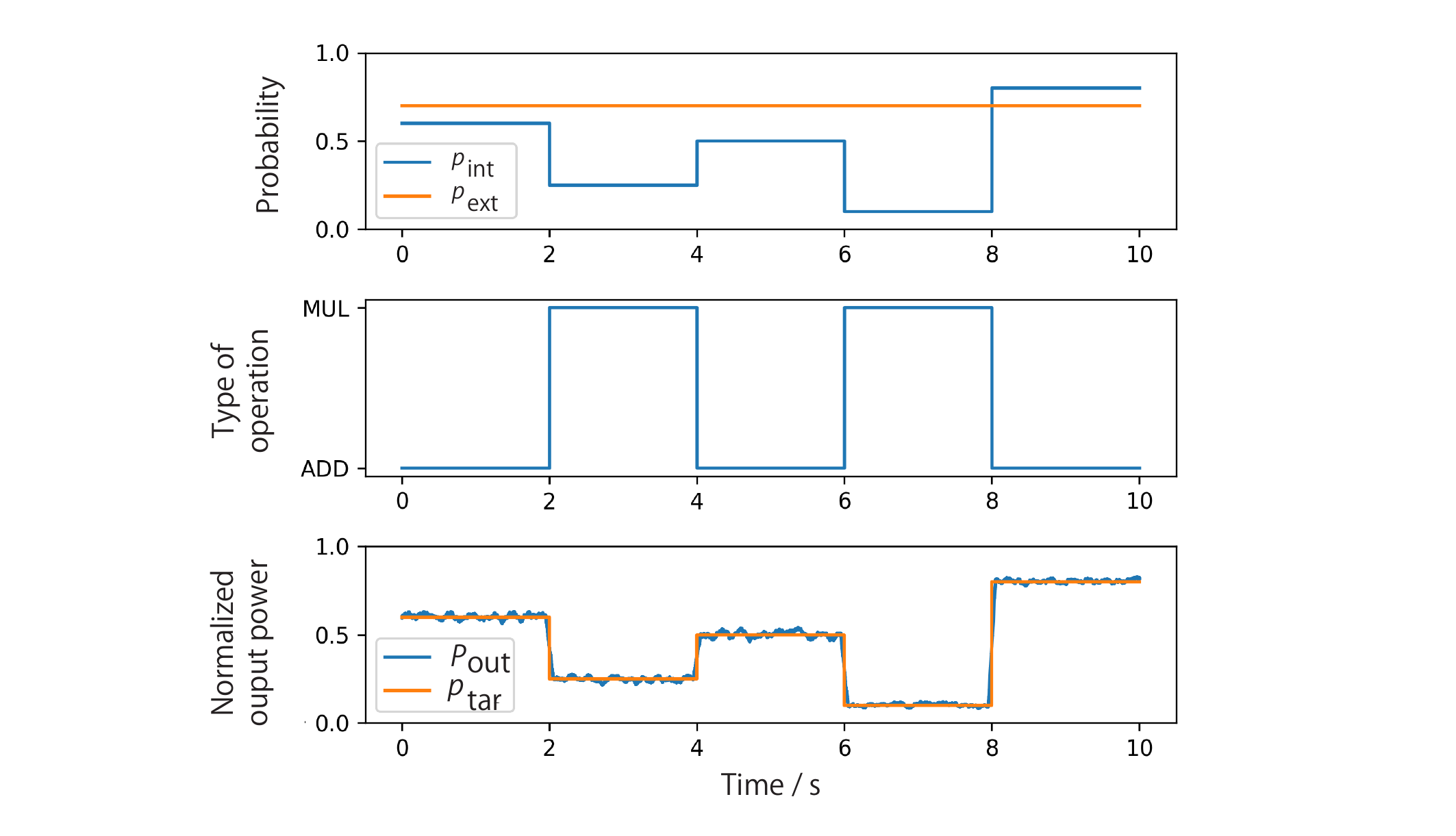}
        \caption{Results of power management based on stochastic power processing. }
        \label{fig:manage2}
    \end{figure}
    
    \subsection{Setups}
    We consider a connected system of two subsystems comprising a source and a load, as shown in Fig.~\ref{fig:manage1}~\textit{a}. 
    The subsystem of interest (subsystem A) can receive power packets from the other subsystem (subsystem B) when it has surplus capacity. 
    Here, the redundancy corresponds to time slots that are not used in subsystem B since we consider packet-based power management. 
    
    Subsequently, for the management, we assume the following.
    \begin{enumerate}
        \item Subsystem A knows the demand of the load represented as a probability (or density) of the power packet supply. 
        \item Subsystem A can regulate the output probability of its source. 
        \item Subsystem B does not have any information on the demand of the load. 
        \item The unused time slots of subsystem B occur at a constant probability. 
    \end{enumerate}
    The management objective is to satisfy the load demand in terms of the supply probability of the power packets.
    We regulate the power packet supply of subsystem A through stochastic power processing and achieve the desired output probability. 
    
    We denote the probability of the internal supply from the source of subsystem A as $p_\mathrm{int}$ and that of the external supply from the source of subsystem B as $p_\mathrm{ext}$. 
    We set a step-like change of the target probability $p_\mathrm{tar}$, which represents the load demand, with an identical holding time interval of $1\,\mathrm{s}$. 
    
    When constant values of $p_\mathrm{ext},p_\mathrm{tar} \in (0,1]$ are given, $p_\mathrm{int}$ is calculated based on the type of operation.
    In the case of multiplication, 
    \begin{equation}
        p_\mathrm{int} = p_\mathrm{tar}/p_\mathrm{ext},
        \label{eq:mul}
    \end{equation}
    where $p_\mathrm{int}$ is defined only when $0 \leq p_\mathrm{tar} \leq p_\mathrm{ext}$ holds. 
    In the case of addition, 
    \begin{equation}
        p_\mathrm{int} = 2 p_\mathrm{tar} - p_\mathrm{ext},
        \label{eq:add}
    \end{equation}
    where $p_\mathrm{int}$ is defined only when $p_\mathrm{ext}/2 \leq p_\mathrm{tar} \leq (1+p_\mathrm{ext})/2$ holds. 
    The in/out relationships described in equations (\ref{eq:mul})~and~(\ref{eq:add}) and their constraints are presented in Fig.~\ref{fig:manage1}~\textit{b} for $p_\mathrm{ext} = 0.7$. 
    The resulting output regions are divided into three: one where both multiplication and addition can realize the target and two where only one of the operations can realize the target. 
    In general, the multiplication and addition operations have wider coverage in the low and high-probability regions, respectively. 
    This can be attributed to the qualitative interpretations that the multiplication thins out one input stream based on the other and that the addition adds up two streams. 
    We switch the two operations based on the target probability. 
    
    \subsection{Results}
    Figure~\ref{fig:manage2} presents the result of power management. 
    Figures~\ref{fig:manage2}~\textit{a}~and~\textit{b} depict the probability of the two power packet supplies and the type of operation applied to them. 
    The variable input probability is regulated based on the target and the type of operation. 
    Figure~\ref{fig:manage2}~\textit{c} depicts the average output power of the router overlayed on the target. 
    The values are both normalized by the same base value used in the previous section, namely the output with $(p_\mathrm{f},p_\mathrm{b}) = (1,1)$. 
    Additionally, a moving average of the time window $100\,\mathrm{ms}$ is applied to the output power. 
    The result indicates that the load demand was successfully satisfied in terms of the average power supply.
    
    \section{Conclusions}
    In this article, we proposed stochastic power processing based on the logic operation of power. 
    The processing scheme is implemented as a function of a power packet router. 
    Power regulation was achieved by performing a logic operation of two power packet streams considering the power supply as a probability of pulse power occurrence. 
    We developed hardware and software configurations and demonstrated the feasibility of the proposed scheme through experimental analysis. 
    Furthermore, we demonstrated its application to power management under an example scenario with a simple but essential configuration of two cooperating subnetworks. 
    
    The stochastic power processing realizes the function of the power packet router as a filter of random input to form a desired output \cite{Wiener-1958,Baek-2021}. 
    This helps in addressing practical problems in the full utilization of energy harvesting. 
    Energy harvesting involves selectively extracting power for a specific purpose from the high-entropy power distributed in the environment \cite{Wang-2019}. 
    This operation is recognized as a control of the spatiotemporal distribution of power \cite{Nawata.etal-2018,Gelenbe.Ceran-2016}. 
    The proposed scheme embodies it as a dynamic function of a network of power packet routers. 
    
    
    
    \section*{acknowledgment}
    This work was partially supported by JST-OPERA Program Grant Number JPMJOP1841, by JSPS KAKENHI Grant Numbers 20H02151 and 20K14732, and by Kayamori Foundation of Informational Science Advancement. 
    
    
    \bibliographystyle{RS} 
    \bibliography{bibliography} 

\end{document}